\documentclass[aps, pre, superscriptaddress, twocolumn]{revtex4}
\usepackage{bm}
\usepackage{amsmath}
\usepackage{amsfonts}
\usepackage{amsthm}
\usepackage{xcolor}
\usepackage[pdftex]{graphicx}
\usepackage[utf8]{inputenc}

\begin{document}

\newcommand{\defeq}{\mathrel{\mathop:}=}

\title{On the possible distributions of temperature in nonequilibrium steady states}

\author{Sergio Davis}
\email{sergio.davis@cchen.cl}

\affiliation{Comisi\'on Chilena de Energía Nuclear, Casilla 188-D, Santiago, Chile}
\affiliation{Departamento de F\'isica, Facultad de Ciencias Exactas, Universidad Andres Bello. Sazi\'e 2212, piso 7, 8370136, Santiago, Chile.}

\date{\today}

\begin{abstract}
Superstatistics is a framework in nonequilibrium statistical mechanics that successfully describes a wide variety of complex systems, including hydrodynamic turbulence, 
weakly-collisional plasmas, cosmic rays, power grid fluctuations, among several others. In this work we analyze the class of nonequilibrium steady-state systems consisting of a 
subsystem and its environment, and where the subsystem is described by the superstatistical framework. In this case we provide an answer to the mechanism by which a broad 
distribution of temperature arises, namely due to correlation between subsystem and environment. We prove that there is a unique microscopic definition $\mathcal{B}$ of 
inverse temperature compatible with superstatistics, in the sense that all moments of $\mathcal{B}$ and $\beta=1/(k_B T)$ coincide. The function $\mathcal{B}$ however, cannot 
depend on the degrees of freedom of the system itself, only on the environment, in full agreement with our previous impossibility theorem [Physica A \textbf{505}, 864-870 (2018)].
The present results also constrain the possible joint ensembles of system and environment compatible with superstatistics.
\end{abstract}

\maketitle

\section{Introduction}

An increasing number of complex systems out of equilibrium are described by non-canonical statistics, most remarkably ensembles described by power laws instead of 
exponential distributions. Currently two theoretical frameworks employed to explain such non-canonical states stand out, namely Tsallis' nonextensive statistics~\cite{Tsallis2009,
Tsallis2011} and superstatistics~\cite{Beck2003,Beck2004,Beck2011,Hanel2011}. While Tsallis statistics postulates a generalization of the Boltzmann-Gibbs entropy, superstatistics aims to 
describe the same steady states and others without the need for redefining the entropy. It has been successfully applied to the description of turbulence~\cite{Reynolds2003,Beck2007}, 
space and laboratory plasmas~\cite{Ourabah2015}, solar flares~\cite{Baiesi2006}, fluctuations in electrical power grids~\cite{Schafer2018}, cosmology~\cite{Jizba2013}, among 
several others~\cite{Beck2005, Briggs2007, GarciaMorales2011, Alves2016}.

Fundamentally, superstatistics replaces the single value of the inverse temperature parameter $\beta=1/(k_B T)$ in the canonical ensemble by a \emph{probability distribution of 
(inverse) temperatures}, that we will express through the Bayesian notation $P(\beta|S)$, where $S$ denotes a steady state. More precisely, we will define superstatistics for a 
system with microstates $\bm x=(x_1, \ldots, x_N)$ through the joint distribution
\begin{align}
P(\bm x, \beta | S) & = P(\bm x|\beta) P(\beta|S) \nonumber \\
                    & = \left[\frac{\exp(-\beta H(\bm x))}{Z(\beta)}\right]\; P(\beta|S),
\label{eq_super_joint}
\end{align}
where $H(\bm x)$ is the Hamiltonian of the system, $Z(\beta)$ the partition function,
\begin{equation}
Z(\beta) = \int d\bm{x} \exp(-\beta H(\bm x)) = \int dE \Omega(E)\exp(-\beta E),
\end{equation}
and $\Omega(E)=\int d\bm{x}\delta(E-H(\bm x))$ the density of states of $H$. The probability density of microstates is then expressed through integration over all possible 
values of $\beta$ as~\cite{Sattin2006}
\begin{equation}
P(\bm x|S) = \int d\beta \left[\frac{\exp(-\beta H(\bm x))}{Z(\beta)}\right] P(\beta|S).
\end{equation}

From this is clear that the probability of the microstate only depends on its energy, and we can write this dependence as $P(\bm x|S)=\rho(H(\bm x))$,
where we have defined the \emph{ensemble function}
\begin{equation}
\rho(E) \defeq \int d\beta \exp(-\beta E)f(\beta),
\end{equation}
and the \emph{superstatistical weight function} 
\begin{equation}
f(\beta) \defeq P(\beta|S)/Z(\beta)
\end{equation}
for convenience.

\section{A microscopic definition of temperature}
\label{sect_microdef}

The fact that superstatistics assumes a statistical distribution of temperature raises several questions, most important of which is the existence and nature of temperature fluctuations~\cite{Sattin2018}. When $\beta$ is broadly distributed and we can assign a variance $\big<(\delta \beta)^2\big>_S$, is it just statistical uncertainty or is 
there a fluctuating quantity?

Although it would be natural to assume the existence of one, or even a family of \emph{microscopic definitions of temperature}, it has been recently shown~\cite{Davis2018}
that no function $B(\bm x)$ can be identified one-to-one with the inverse temperature $\beta$~\cite{Note1} in superstatistics. In other words, for a superstatistical ensemble $S$, 
no function $B(\bm x)$ can be constructed such that
\begin{equation}
\Big<g(\beta)\Big>_S = \Big<g(B)\Big>_S,
\end{equation}
for an arbitrary function $g(\beta)$. If such a function $B$ existed, we would be able to measure $\beta$ and its statistical properties (e.g. all its moments)
\emph{from within the system}, by instead collecting statistics of $B$. In particular, a histogram of $B$ would converge to the superstatistical distribution $P(\beta|S)$.

However, there is still hope. In this work it is shown that by considering an extended setup of system $\bm x$ and environment $\bm y$ with joint probability
\begin{equation}
P(\bm x, \bm y|S) = p(H(\bm x), G(\bm y)),
\label{eq_joint_model}
\end{equation}
then \emph{it is possible} to define a function $\mathcal{B}$ such that
\begin{equation}
\Big<g(\beta)\Big>_S = \Big<g(\mathcal{B})\Big>_S
\label{eq_condition}
\end{equation}
for any $g(\beta)$. Moreover, the function $\mathcal{B}$ only depends on the environment $\bm y$, and is uniquely defined by
\begin{equation}
\mathcal{B} = -\frac{\partial}{\partial E}\ln p(E, G) = \mathcal{B}(G).
\label{eq_solution}
\end{equation}

The fact that $\mathcal{B}(G)$ cannot depend on $\bm x$, directly implies an even stronger version of the original impossibility theorem of Ref.~\cite{Davis2018}, that also 
rules out the definitions $B(\bm x)$ which are dependent of the ensemble function $\rho(E)$. Note also that, if the number of degrees of freedom of $\bm{y}$ is large enough (i.e. the
system $\bm{x}$ is in contact with an infinite reservoir), by the asymptotic equipartition property (AEP)~\cite{CoverThomas2006} the fluctuations of $G$ vanish and 
$\mathcal{B}(G)$ becomes constant, hence the system $\bm{x}$ approaches the canonical ensemble.

From integration of Eq. \ref{eq_solution} and proper normalization, we see that the only ensembles leading to superstatistics for $\bm{x}$ are of the form
\begin{align}
P(\bm{x}, \bm{y}|S) & = \rho_G(G(\bm y))\left[\frac{\exp(-\beta H(\bm x))}{Z(\beta)}\right]\Big|_{\beta=\mathcal{B}(G(\bm y))} \nonumber \\
                    & = P(\bm y|S)\times P(\bm x|\beta = \mathcal{B}(G(\bm y))),
\label{eq_joint_final}
\end{align}
which directly implies that
\begin{equation}
P(\bm x|\bm y, S) = P(\bm x|\beta = \mathcal{B}(G(\bm y))).
\label{eq_defsuper}
\end{equation}

In fact, Eq. \ref{eq_defsuper} can be understood as the defining condition for superstatistics, namely that the statistical ensemble of the system $\bm{x}$ under a ``frozen'' 
environment $\bm{y}$ must be completely equivalent to a canonical ensemble with inverse temperature $\mathcal{B}(G(\bm y))$. In other words, the system and its environment 
are coupled through the fluctuating inverse temperature $\mathcal{B}$, as $P(\bm x, \bm y|S)$ is only separable into a product $P(\bm x|S) \times P(\bm y|S)$ when $\mathcal{B}$ 
is a constant (the canonical ensemble). We arrive at the following conclusion: \emph{correlation between system and environment implies non-canonical superstatistics}.

Probably the most important case in which superstatistics is obtained in this way, is the case with
\begin{equation}
P(\bm x, \bm y|S) = p(H(\bm x)+G(\bm y)),
\end{equation}
and $H \ll G$. We have, by expanding $\ln p(H+G)$ to first order around $H=0$, that
\begin{align}
P(\bm x, \bm y|S) & = \exp(\ln p(H(\bm x)+G(\bm y)))  \nonumber \\
                    & \approx p(G(\bm y))\exp(-\mathcal{B}(G(\bm y))H(\bm y)) \nonumber \\
                    & \approx \rho_G(G(\bm y))\left[\frac{\exp(-\mathcal{B}(G(\bm y))H(\bm y))}{Z(\mathcal{B}(G(\bm y)))}\right],
\end{align}
which is Eq. \ref{eq_joint_final}, with
\begin{equation}
\mathcal{B}(G) = -\frac{\partial}{\partial G}\ln p(G).
\end{equation}

That is, in this case the temperature function matches the fundamental temperature~\cite{Davis2019} of the whole (system plus environment), evaluated at the energy of
the environment. This is precisely the result recently shown in Ref.~\cite{Davis2019b} in the context of collisionless plasmas, but now understood in a more general way. This 
case is also connected with superstatistics in small, correlated systems, see for instance Ref.~\cite{Dixit2013}.

\section{Some new properties of temperature in superstatistics}

One useful consequence of Eq. \ref{eq_defsuper} is
\begin{align}
P(\bm x|G, S) & = \int d\bm{y}P(\bm x|\bm y, S) P(\bm y|G, S) \nonumber \\
              & = \int d\bm{y}P(\bm x|\beta=\mathcal{B}(G(\bm y)))\left[\frac{\delta(G(\bm y)-G)}{W(G)}\right]\nonumber \\
              & = P(\bm x|\beta=\mathcal{B}(G)),
\label{eq_canon_G}
\end{align}
where $W(G) = \int d\bm{y}\delta(G(\bm y)-G)$ is the density of states of $G$. Essentially what Eqs. \ref{eq_defsuper} and \ref{eq_canon_G} tell us is that 
the states of knowledge $(\bm y, S)$ and $(G, S)$ regarding features of the system $\bm{x}$ can be replaced by canonical states,
\begin{align}
\big<\cdot\big>_{\bm y, S} & \rightarrow \big<\cdot\big>_{\beta=\mathcal{B}(G(\bm y))}, \\
\big<\cdot\big>_{G, S} & \rightarrow \big<\cdot\big>_{\beta=\mathcal{B}(G)}.
\end{align}

For instance, the conditional distribution of energy given $G$ is directly $P(E|\beta=\mathcal{B}(G))$. Using these rules we can further validate our identification of 
$\mathcal{B}$ with the thermodynamical inverse temperature of the system by showing that
\begin{equation}
\Big<\beta_\Omega\Big>_{G, S} = \Big<\hat{\beta}\Big>_{G, S} = \mathcal{B}(G),
\label{eq_equalities}
\end{equation}
where 
\begin{equation}
\beta_\Omega(E)=\frac{\partial}{\partial E}\ln \Omega(E)
\end{equation}
is the \textit{microcanonical inverse temperature} and $\hat{\beta}$ is the so-called \textit{dynamical temperature} estimator,
\begin{equation}
\hat{\beta}(\bm x) = \nabla\cdot \left[\frac{\bm{\omega}(\bm x)}{\bm{\omega}(\bm x)\cdot \nabla H}\right],
\end{equation}
where $\nabla=\partial/\partial \bm{x}$ and the identity
\begin{equation}
\Big<\hat{\beta}\Big>_E = \beta_\Omega(E),
\label{eq_betahat}
\end{equation}
holds~\cite{Rugh1997,Rickayzen2001}. The proof of Eq. \ref{eq_equalities} uses the general property~\cite{Note2} 
\begin{equation}
\Big<\frac{\partial}{\partial x} \ln P(x|\mathcal{I})\Big>_\mathcal{I} = 0
\end{equation}
for any random variable $x$ and state of knowledge $\mathcal{I}$. For the canonical energy distribution we have
\begin{equation}
\Big<\frac{\partial}{\partial E}\ln P(E|\beta)\Big>_\beta = 0 = -\beta + \Big<\beta_\Omega\Big>_\beta,
\end{equation}
which, together with Eq. \ref{eq_betahat}, allows us to write
\begin{align}
\Big<\hat{\beta}\Big>_\beta & = \int dE P(E|\beta)\Big<\hat{\beta}\Big>_E \nonumber \\
                            & = \int dE P(E|\beta)\beta_\Omega(E) = \Big<\beta_\Omega\Big>_\beta.
\end{align}

\noindent
Therefore, we can write
\begin{equation}
\Big<\beta_\Omega\Big>_\beta = \Big<\hat{\beta}\Big>_\beta = \beta,
\end{equation}
which after replacing $\big<\cdot\big>_{G, S} \Longleftrightarrow \big<\cdot\big>_{\beta=\mathcal{B}(G)}$ yields Eq. \ref{eq_equalities}. Another 
interesting consequence is the connection between $\mathcal{B}(G)$ and the conditional expectation of energy given $G$,
\begin{equation}
\big<H\big>_{G,S} = \big<H\big>_{\beta=\mathcal{B}(G)} = \left[-\frac{\partial}{\partial \beta}\ln Z(\beta)\right]\Big|_{\beta=\mathcal{B}(G)}.
\label{eq_H_GS}
\end{equation}
 
This gives a more intuitive meaning to the superstatistical (inverse) temperature $\mathcal{B}(G)$ as the conjugate quantity (in the thermodynamical sense) 
to $\big<H\big>_{S, G}$. 

\section{Inverse temperature fluctuations}

Despite the fact that the inverse temperature $\mathcal{B}(G)$ cannot be accessed from the system $\bm x$, its variance (and therefore, that of $\beta$) can be computed from within the 
system, through the microcanonical inverse temperature $\beta_\Omega(E)$, and is given by
\begin{equation}
\Big<(\delta \mathcal{B})^2\Big>_S = \Big<(\delta \beta_\Omega)^2\Big>_S + \Big<\frac{\partial \beta_\Omega}{\partial E}\Big>_S.
\label{eq_fluct}
\end{equation}

\noindent 
To prove this assertion, we first compute $P(E, G|S)$,
\begin{align}
P(E, G|S) & = P(E|G, S)\times P(G|S) \nonumber \\
          & = P(E|\beta=\mathcal{B}(G))\times P(G|S) \nonumber \\
          & = \left[\frac{\exp(-\mathcal{B}(G)E)}{Z(\mathcal{B}(G))}\Omega(E)\right] \times P(G|S),
\end{align}
and then, from the conjugate variables theorem~\cite{Davis2012}, obtain the identity
\begin{align}
\Big<\frac{\partial \omega}{\partial E}\Big>_S & = -\Big<\omega\frac{\partial}{\partial E}\ln P(E, G|S)\Big>_S \nonumber \\
      & = \Big<\omega\Big(\mathcal{B}(G)-\beta_\Omega(E)\Big)\Big>_S,
\end{align}
which is valid for any function $\omega=\omega(E, G)$ such that $\partial \omega/\partial E$ exists. Using the constant function $\omega_1(E, G)=1$ we confirm that 
\begin{equation}
\Big<\mathcal{B}\Big>_S = \Big<\beta_\Omega\Big>_S,
\label{eq_w1}
\end{equation}
while the choices $\omega_2(E, G)=\mathcal{B}(G)-\beta_\Omega(E)$ and $\omega_3(E, G)=\mathcal{B}(G)$ produce

\begin{equation}
2\Big<\mathcal{B}\beta_\Omega\Big>_S = \Big<\mathcal{B}^2\Big>_S + \Big<\beta_\Omega^2\Big>_S + \Big<\frac{\partial \beta_\Omega}{\partial E}\Big>_S,
\label{eq_w2}
\end{equation}
and
\begin{equation}
\Big<\mathcal{B}^2\Big>_S = \Big<\mathcal{B}\beta_\Omega\Big>_S,
\label{eq_w3}
\end{equation}
respectively. Eq. \ref{eq_fluct} readily follows by combining Eqs. \ref{eq_w1}, \ref{eq_w2} and \ref{eq_w3}.

\section{Proof of the unique definition of temperature}

Now the proof of our main result, that Eq. \ref{eq_solution} is the unique solution of the condition given in Eq. \ref{eq_condition}, is given. 
First we establish that $\mathcal{B}(\bm x, \bm y)$ can only depend on $\bm x$ through $H(\bm x)$. In order to see why this is so, 
consider the joint distribution of $\bm{x}$ and $\beta$ in Eq. \ref{eq_super_joint},
\begin{equation}
P(\bm x, \beta|S) = \exp(-\beta H(\bm x))f(\beta),
\end{equation}
and impose that $P(\bm x, \beta|S) = P(\bm x, \mathcal{B}=\beta|S)$. We have
\begin{equation}
\exp(-\beta E)f(\beta) = \int d\bm{y} p(E, G(\bm y))\delta(\mathcal{B}-\beta),
\label{eq_expbetae_f}
\end{equation}
with $E \defeq H(\bm x)$. Multiplying by $\exp(\beta E)$ and using the properties of the Dirac delta to replace $\exp(\beta E)$ by $\exp(\mathcal{B}E)$ inside the
integral on the right-hand side, we obtain
\begin{equation}
f(\beta) = \int d\bm{y} p(E, G(\bm y))\exp(\mathcal{B}E)\delta(\mathcal{B}-\beta),
\end{equation}
hence the integral cannot depend on $\bm x$ at all. This is only possible if either $\mathcal{B}$ does not depend on $\bm x$, or if it depends on $\bm x$ through $H(\bm x)$.
Both cases can be condensed into the form $\mathcal{B}(\bm x, \bm y)=\mathcal{B}(H(\bm x), \bm y)$, and we can write Eq. \ref{eq_expbetae_f} as
\begin{equation}
\exp(-\beta E)f(\beta) = \int d\bm{y} p(E, G(\bm y))\delta(\mathcal{B}(E, \bm y)-\beta).
\label{eq_expbeta}
\end{equation}

\noindent
Applying $\partial^n/\partial E^n$ on both sides, we have
\begin{equation}
\exp(-\beta E)f(\beta)(-\beta)^n = \int d\bm{y} \frac{\partial^n}{\partial E^n}\Big(p \cdot \delta(\mathcal{B}-\beta)\Big)
\label{eq_partial_n}
\end{equation}
and now multiplying by $\Omega(E)$ and integrating in both $E$ and $\beta$, we have for the left-hand side
\begin{align}
\int d\beta & \left[\int dE\;\frac{\exp(-\beta E) \Omega(E)}{Z(\beta)}\right] (-\beta)^n P(\beta|S) \nonumber \\
  & = \int d\beta (-\beta)^n P(\beta|S) = \Big<(-\beta)^n\Big>_S,
\end{align}
while the right-hand side becomes
\begin{align}
\int & dE \Omega(E) \int d\bm{y} \frac{\partial^n}{\partial E^n}\Big(p \cdot \int d\beta \delta(\mathcal{B}-\beta)\Big) \nonumber \\
     & = \int d\bm{x}d\bm{y} p(H, G)\left[\frac{1}{p}\frac{\partial^n p}{\partial E^n}\right]\Big|_{H,G}\hspace{-6pt}
     = \left<\left[\frac{1}{p}\frac{\partial^n p}{\partial E^n}\right]\right>_S.
\label{eq_beta_moments}
\end{align}

We have obtained then a relationship between the moments of $\beta$ and the derivatives of the ensemble function $p(E, G)$,
\begin{equation}
\Big<\beta^n\Big>_S = \left<\frac{(-1)^n}{p}\left[\frac{\partial^n p}{\partial E^n}\right]\right>_S = \Big<\mathcal{B}^n\Big>_S,
\label{eq_power_B}
\end{equation}
where the last equality is imposed by Eq. \ref{eq_condition}. From this we extract two conclusions. First, $\mathcal{B}(E, \bm y)=\mathcal{B}(E, G(\bm y))$,
and second, $\mathcal{B}$ is only a functional of the joint ensemble function $p$, and does not depend on the densities of states $\Omega(E)$ and $W(G)$.
We can at this point already recognize $\mathcal{B}$ by using $n=1$ in Eq. \ref{eq_power_B},
\begin{equation}
\mathcal{B}= -\frac{\partial}{\partial E}\ln p(E, G),
\end{equation}
however, we can aim for a more exhaustive proof. Using the fact that
\begin{equation}
-\frac{\partial}{\partial E}\ln \Big[\exp(-\beta E)f(\beta)\Big] = \beta,
\end{equation}
and replacing Eq. \ref{eq_expbeta}, we obtain
\begin{equation}
\beta = -\frac{\int dG W(G)\frac{\partial}{\partial E}\left[p(E, G) \delta(\mathcal{B}(E, G) - \beta)\right]}{\int dG W(G) p(E, G) \delta(\mathcal{B}(E, G) - \beta)},
\end{equation}
where we have introduced the density of states $W(G)$ on both integrals. This equation can be rearranged as a functional of $W$ which is identically zero,
\begin{equation}
\int dG W(G) \left\{-\frac{\partial}{\partial E}\Big(p \delta(\mathcal{B} - \beta)\Big) - p\delta(\mathcal{B}-\beta)\beta\right\}=0.
\end{equation}

As neither $p$ nor $\mathcal{B}$ depend on $W$, we have
\begin{equation}
-\frac{\partial}{\partial E}\Big(p(E, G) \delta(\mathcal{B} - \beta)\Big) = p(E, G)\delta(\mathcal{B}-\beta)\beta,
\label{eq_functional}
\end{equation}
which by integrating $\beta$ on both sides, becomes
\begin{equation}
-\frac{\partial}{\partial E} p(E, G) = p(E, G)\mathcal{B},
\end{equation}
immediately yielding a unique definition of $\mathcal{B}$,
\begin{equation}
\mathcal{B} = -\frac{\partial}{\partial E}\ln p(E, G).
\label{eq_beta_def1}
\end{equation}

\noindent
Replacing this value of $\mathcal{B}$ into Eq. \ref{eq_functional}, we arrive at
\begin{equation}
-p(E, G)\frac{\partial}{\partial E}\delta(\mathcal{B} - \beta) = 0,
\end{equation}
hence $\mathcal{B}(E, G)$ does not depend on $E$, and
\begin{equation}
\frac{\partial \mathcal{B}}{\partial E} = -\frac{\partial^2}{\partial E^2}\ln p(E, G) = 0.
\label{eq_second_der}
\end{equation}

Eqs. \ref{eq_beta_def1} and \ref{eq_second_der} together lead to the solution in Eq. \ref{eq_solution}.

\noindent
We have then, by integrating Eq. \ref{eq_second_der} twice, that
\begin{equation}
p(E, G) = p_0(G)\exp(-\mathcal{B}(G)E),
\label{eq_joint_ensemble}
\end{equation}
where the function $p_0(G)$ remains to be determined. Integrating $E$ from $P(E, G|S)$ we have 
\begin{equation}
P(G|S) = \rho_G(G)W(G) = W(G)\int dE \Omega(E)p(E, G),
\end{equation}
thus replacing Eq. \ref{eq_joint_ensemble} gives
\begin{align}
\rho_G(G) & = p_0(G)\int dE\Omega(E)\exp(-\mathcal{B}(G)E) \nonumber \\
          & = p_0(G)Z(\mathcal{B}(G)).
\label{eq_rhoG}
\end{align}

\noindent 
Replacing Eq. \ref{eq_rhoG} into Eq. \ref{eq_joint_ensemble} we can finally write
\begin{equation}
p(E, G) = \rho_G(G)\left[\frac{\exp(-\mathcal{B}(G)E)}{Z(\mathcal{B}(G))}\right],
\label{eq_joint_ensemble_final}
\end{equation}
which gives the full expression for $P(\bm x, \bm y|S)$ in Section \ref{sect_microdef} (Eq. \ref{eq_joint_final}). Using this expression for the joint 
ensemble function $p(E, G)$ is straightforward to verify that
\begin{equation}
\frac{(-1)^n}{p}\frac{\partial^n p}{\partial E^n} = \mathcal{B}(G)^n,
\end{equation}
as required by Eq. \ref{eq_power_B}. The proof of Eq. \ref{eq_condition} follows by introducing the series expansion of an arbitrary function 
$g(\beta)=\sum_{n=0}^{\infty} C_n\beta^n$ and taking expectation in $S$, which yields
\begin{equation}
\big<g(\beta)\big>_S = \sum_{n=0}^\infty C_n\big<\beta^n\big>_S = \sum_{n=0}^\infty C_n\big<\mathcal{B}^n\big>_S = \big<g(\mathcal{B})\big>_S.
\end{equation}

\section{Concluding remarks}

We have proved that there is a unique microscopic definition of inverse temperature (Eq. \ref{eq_solution}) fully compatible with superstatistics, in the sense that the 
parameter $\beta$ can be replaced anywhere by a function $\mathcal{B}(G)$ of the environment. The fact that $\mathcal{B}$ cannot depend on $\bm x$ seems to rule out a spatial 
distribution of temperatures in the superstatistical framework. However, it is still possible to maintain a frequentist interpretation of superstatistics with a \textit{fluctuating,    
instantaneous (inverse) temperature} $\mathcal{B}(G)$, which is a global property \emph{of the environment}. We are led to the conclusion that temperature fluctuations essentially 
map the fluctuations of the energy of the environment, and it is precisely this fluctuating temperature that correlates the system and its environment.

\section*{Acknowledgments}

The author gratefully acknowledges support by Anillo ACT-172101 grant, and partial support from FONDECYT 1171127 grant.


\begin{thebibliography}{28}
\expandafter\ifx\csname natexlab\endcsname\relax\def\natexlab#1{#1}\fi
\expandafter\ifx\csname bibnamefont\endcsname\relax
  \def\bibnamefont#1{#1}\fi
\expandafter\ifx\csname bibfnamefont\endcsname\relax
  \def\bibfnamefont#1{#1}\fi
\expandafter\ifx\csname citenamefont\endcsname\relax
  \def\citenamefont#1{#1}\fi
\expandafter\ifx\csname url\endcsname\relax
  \def\url#1{\texttt{#1}}\fi
\expandafter\ifx\csname urlprefix\endcsname\relax\def\urlprefix{URL }\fi
\providecommand{\bibinfo}[2]{#2}
\providecommand{\eprint}[2][]{\url{#2}}

\bibitem[{\citenamefont{Tsallis}(2009)}]{Tsallis2009}
\bibinfo{author}{\bibfnamefont{C.}~\bibnamefont{Tsallis}},
  \emph{\bibinfo{title}{Introduction to nonextensive statistical mechanics:
  approaching a complex world}} (\bibinfo{publisher}{Springer Science \&
  Business Media}, \bibinfo{year}{2009}).

\bibitem[{\citenamefont{Tsallis}(2011)}]{Tsallis2011}
\bibinfo{author}{\bibfnamefont{C.}~\bibnamefont{Tsallis}},
  \bibinfo{journal}{Entropy} \textbf{\bibinfo{volume}{13}},
  \bibinfo{pages}{1765} (\bibinfo{year}{2011}).

\bibitem[{\citenamefont{Beck and Cohen}(2003)}]{Beck2003}
\bibinfo{author}{\bibfnamefont{C.}~\bibnamefont{Beck}} \bibnamefont{and}
  \bibinfo{author}{\bibfnamefont{E.}~\bibnamefont{Cohen}},
  \bibinfo{journal}{Phys. A} \textbf{\bibinfo{volume}{322}},
  \bibinfo{pages}{267} (\bibinfo{year}{2003}).

\bibitem[{\citenamefont{Beck}(2004)}]{Beck2004}
\bibinfo{author}{\bibfnamefont{C.}~\bibnamefont{Beck}}, \bibinfo{journal}{Cont.
  Mech. Thermodyn.} \textbf{\bibinfo{volume}{16}}, \bibinfo{pages}{293}
  (\bibinfo{year}{2004}).

\bibitem[{\citenamefont{Beck}(2011)}]{Beck2011}
\bibinfo{author}{\bibfnamefont{C.}~\bibnamefont{Beck}}, \bibinfo{journal}{Phil.
  Trans. R. Soc. A} \textbf{\bibinfo{volume}{369}}, \bibinfo{pages}{453}
  (\bibinfo{year}{2011}).

\bibitem[{\citenamefont{Hanel et~al.}(2011)\citenamefont{Hanel, Thurner, and
  Gell-Mann}}]{Hanel2011}
\bibinfo{author}{\bibfnamefont{R.}~\bibnamefont{Hanel}},
  \bibinfo{author}{\bibfnamefont{S.}~\bibnamefont{Thurner}}, \bibnamefont{and}
  \bibinfo{author}{\bibfnamefont{M.}~\bibnamefont{Gell-Mann}},
  \bibinfo{journal}{Proc. Nat. Acad. Sci.} \textbf{\bibinfo{volume}{108}},
  \bibinfo{pages}{6390} (\bibinfo{year}{2011}).

\bibitem[{\citenamefont{Reynolds}(2003)}]{Reynolds2003}
\bibinfo{author}{\bibfnamefont{A.}~\bibnamefont{Reynolds}},
  \bibinfo{journal}{Phys. Rev. Lett.} \textbf{\bibinfo{volume}{91}},
  \bibinfo{pages}{084503} (\bibinfo{year}{2003}).

\bibitem[{\citenamefont{Beck}(2007)}]{Beck2007}
\bibinfo{author}{\bibfnamefont{C.}~\bibnamefont{Beck}}, \bibinfo{journal}{Phys.
  Rev. Lett.} \textbf{\bibinfo{volume}{98}}, \bibinfo{pages}{064502}
  (\bibinfo{year}{2007}).

\bibitem[{\citenamefont{Ourabah et~al.}(2015)\citenamefont{Ourabah, Gougam, and
  Tribeche}}]{Ourabah2015}
\bibinfo{author}{\bibfnamefont{K.}~\bibnamefont{Ourabah}},
  \bibinfo{author}{\bibfnamefont{L.~A.} \bibnamefont{Gougam}},
  \bibnamefont{and} \bibinfo{author}{\bibfnamefont{M.}~\bibnamefont{Tribeche}},
  \bibinfo{journal}{Phys. Rev. E} \textbf{\bibinfo{volume}{91}},
  \bibinfo{pages}{12133} (\bibinfo{year}{2015}).

\bibitem[{\citenamefont{Baiesi et~al.}(2006)\citenamefont{Baiesi, Paczuski, and
  Stella}}]{Baiesi2006}
\bibinfo{author}{\bibfnamefont{M.}~\bibnamefont{Baiesi}},
  \bibinfo{author}{\bibfnamefont{M.}~\bibnamefont{Paczuski}}, \bibnamefont{and}
  \bibinfo{author}{\bibfnamefont{A.~L.} \bibnamefont{Stella}},
  \bibinfo{journal}{Phys. Rev. Lett.} \textbf{\bibinfo{volume}{96}},
  \bibinfo{pages}{051103} (\bibinfo{year}{2006}).

\bibitem[{\citenamefont{Schäfer et~al.}(2018)\citenamefont{Schäfer, Beck,
  Aihara, Witthaut, and Timme}}]{Schafer2018}
\bibinfo{author}{\bibfnamefont{B.}~\bibnamefont{Schäfer}},
  \bibinfo{author}{\bibfnamefont{C.}~\bibnamefont{Beck}},
  \bibinfo{author}{\bibfnamefont{K.}~\bibnamefont{Aihara}},
  \bibinfo{author}{\bibfnamefont{D.}~\bibnamefont{Witthaut}}, \bibnamefont{and}
  \bibinfo{author}{\bibfnamefont{M.}~\bibnamefont{Timme}},
  \bibinfo{journal}{Nature Energy} \textbf{\bibinfo{volume}{3}},
  \bibinfo{pages}{119} (\bibinfo{year}{2018}).

\bibitem[{\citenamefont{Jizba and Scardigli}(2013)}]{Jizba2013}
\bibinfo{author}{\bibfnamefont{P.}~\bibnamefont{Jizba}} \bibnamefont{and}
  \bibinfo{author}{\bibfnamefont{F.}~\bibnamefont{Scardigli}},
  \bibinfo{journal}{The European Physical Journal C}
  \textbf{\bibinfo{volume}{73}}, \bibinfo{pages}{2491} (\bibinfo{year}{2013}).

\bibitem[{\citenamefont{Beck et~al.}(2005)\citenamefont{Beck, Cohen, and
  Swinney}}]{Beck2005}
\bibinfo{author}{\bibfnamefont{C.}~\bibnamefont{Beck}},
  \bibinfo{author}{\bibfnamefont{E.~G.} \bibnamefont{Cohen}}, \bibnamefont{and}
  \bibinfo{author}{\bibfnamefont{H.~L.} \bibnamefont{Swinney}},
  \bibinfo{journal}{Phys. Rev. E} \textbf{\bibinfo{volume}{72}},
  \bibinfo{pages}{056133} (\bibinfo{year}{2005}).

\bibitem[{\citenamefont{Briggs and Beck}(2007)}]{Briggs2007}
\bibinfo{author}{\bibfnamefont{K.}~\bibnamefont{Briggs}} \bibnamefont{and}
  \bibinfo{author}{\bibfnamefont{C.}~\bibnamefont{Beck}},
  \bibinfo{journal}{Phys. A} \textbf{\bibinfo{volume}{378}},
  \bibinfo{pages}{498} (\bibinfo{year}{2007}).

\bibitem[{\citenamefont{García-Morales and
  Krischer}(2011)}]{GarciaMorales2011}
\bibinfo{author}{\bibfnamefont{V.}~\bibnamefont{García-Morales}}
  \bibnamefont{and} \bibinfo{author}{\bibfnamefont{K.}~\bibnamefont{Krischer}},
  \bibinfo{journal}{PNAS} \textbf{\bibinfo{volume}{108}},
  \bibinfo{pages}{19535} (\bibinfo{year}{2011}).

\bibitem[{\citenamefont{Alves and Frigori}(2016)}]{Alves2016}
\bibinfo{author}{\bibfnamefont{N.~A.} \bibnamefont{Alves}} \bibnamefont{and}
  \bibinfo{author}{\bibfnamefont{R.~B.} \bibnamefont{Frigori}},
  \bibinfo{journal}{Phys. A} \textbf{\bibinfo{volume}{446}},
  \bibinfo{pages}{195} (\bibinfo{year}{2016}).

\bibitem[{\citenamefont{Sattin}(2006)}]{Sattin2006}
\bibinfo{author}{\bibfnamefont{F.}~\bibnamefont{Sattin}},
  \bibinfo{journal}{Eur. Phys. J. B} \textbf{\bibinfo{volume}{49}},
  \bibinfo{pages}{219} (\bibinfo{year}{2006}).

\bibitem[{\citenamefont{Sattin}(2018)}]{Sattin2018}
\bibinfo{author}{\bibfnamefont{F.}~\bibnamefont{Sattin}},
  \bibinfo{journal}{Phys. Lett. A} \textbf{\bibinfo{volume}{382}},
  \bibinfo{pages}{2551} (\bibinfo{year}{2018}).

\bibitem[{\citenamefont{Davis and Gutiérrez}(2018)}]{Davis2018}
\bibinfo{author}{\bibfnamefont{S.}~\bibnamefont{Davis}} \bibnamefont{and}
  \bibinfo{author}{\bibfnamefont{G.}~\bibnamefont{Gutiérrez}},
  \bibinfo{journal}{Phys. A} \textbf{\bibinfo{volume}{505}},
  \bibinfo{pages}{864} (\bibinfo{year}{2018}).

\bibitem[{Not({\natexlab{a}})}]{Note1}
\bibinfo{note}{The original proof only ruled out functions $B(\bm x)$ which are
  independent of the ensemble function $\rho$. As described in the main text,
  the current derivation is stronger, ruling out all functions $B(\bm x)$
  except the constant function}.

\bibitem[{\citenamefont{Cover and Thomas}(2006)}]{CoverThomas2006}
\bibinfo{author}{\bibfnamefont{T.~M.} \bibnamefont{Cover}} \bibnamefont{and}
  \bibinfo{author}{\bibfnamefont{J.~A.} \bibnamefont{Thomas}},
  \emph{\bibinfo{title}{Elements of Information Theory}}
  (\bibinfo{publisher}{John Wiley and Sons}, \bibinfo{year}{2006}).

\bibitem[{\citenamefont{Davis and Gutiérrez}(2019)}]{Davis2019}
\bibinfo{author}{\bibfnamefont{S.}~\bibnamefont{Davis}} \bibnamefont{and}
  \bibinfo{author}{\bibfnamefont{G.}~\bibnamefont{Gutiérrez}},
  \bibinfo{journal}{Phys. A} \textbf{\bibinfo{volume}{533}},
  \bibinfo{pages}{122031} (\bibinfo{year}{2019}).

\bibitem[{\citenamefont{Davis et~al.}(2019)\citenamefont{Davis, Avaria, Bora,
  Jain, Moreno, Pavez, and Soto}}]{Davis2019b}
\bibinfo{author}{\bibfnamefont{S.}~\bibnamefont{Davis}},
  \bibinfo{author}{\bibfnamefont{G.}~\bibnamefont{Avaria}},
  \bibinfo{author}{\bibfnamefont{B.}~\bibnamefont{Bora}},
  \bibinfo{author}{\bibfnamefont{J.}~\bibnamefont{Jain}},
  \bibinfo{author}{\bibfnamefont{J.}~\bibnamefont{Moreno}},
  \bibinfo{author}{\bibfnamefont{C.}~\bibnamefont{Pavez}}, \bibnamefont{and}
  \bibinfo{author}{\bibfnamefont{L.}~\bibnamefont{Soto}}
  (\bibinfo{year}{2019}), \eprint{arXiv:cond-mat/1906.08072}.

\bibitem[{\citenamefont{Dixit}(2013)}]{Dixit2013}
\bibinfo{author}{\bibfnamefont{P.~D.} \bibnamefont{Dixit}},
  \bibinfo{journal}{J. Chem. Phys.} \textbf{\bibinfo{volume}{138}},
  \bibinfo{pages}{184111} (\bibinfo{year}{2013}).

\bibitem[{\citenamefont{Rugh}(1997)}]{Rugh1997}
\bibinfo{author}{\bibfnamefont{H.~H.} \bibnamefont{Rugh}},
  \bibinfo{journal}{Phys. Rev. Lett.} \textbf{\bibinfo{volume}{78}},
  \bibinfo{pages}{772} (\bibinfo{year}{1997}).

\bibitem[{\citenamefont{Rickayzen and Powles}(2001)}]{Rickayzen2001}
\bibinfo{author}{\bibfnamefont{G.}~\bibnamefont{Rickayzen}} \bibnamefont{and}
  \bibinfo{author}{\bibfnamefont{J.~G.} \bibnamefont{Powles}},
  \bibinfo{journal}{J. Chem. Phys.} \textbf{\bibinfo{volume}{114}},
  \bibinfo{pages}{4333} (\bibinfo{year}{2001}).

\bibitem[{Not({\natexlab{b}})}]{Note2}
\bibinfo{note}{This is true provided the distribution $P(x|\mathcal{I})$
  vanishes at its boundaries, which is the case for the canonical distribution
  of energy with monotonically increasing $\Omega(E)$}.

\bibitem[{\citenamefont{Davis and Guti\'errez}(2012)}]{Davis2012}
\bibinfo{author}{\bibfnamefont{S.}~\bibnamefont{Davis}} \bibnamefont{and}
  \bibinfo{author}{\bibfnamefont{G.}~\bibnamefont{Guti\'errez}},
  \bibinfo{journal}{Phys. Rev. E} \textbf{\bibinfo{volume}{86}},
  \bibinfo{pages}{051136} (\bibinfo{year}{2012}).

\end{thebibliography}

\end{document}